\documentclass[%
reprint,
superscriptaddress,
nofootinbib,
amsmath,amssymb,
aps,
]{revtex4-1}
\usepackage{multirow}
\usepackage{graphicx}
\usepackage{dcolumn}
\usepackage{bm}
\usepackage{color}
\usepackage{afterpage}
\usepackage{url}
\usepackage{longtable}
\usepackage{float}
\newcommand{\icm}{$\text{cm}^{-1}$ }
\newcommand{\celsius}{$^{\circ}$C }

\begin{document}
	
	\preprint{APS/123-QED}
	
	\title{Laser spectroscopy of the 1001\,nm ground state transition in dysprosium}
	
	\author{D. Studer}
	\author{L. Maske}
	\author{P. Windpassinger}
	\author{K. Wendt}
	\affiliation{Institut f\"ur Physik, Johannes Gutenberg-Universit\"at Mainz, 55128 Mainz, Germany}
	
	\date{\today}
	
	\begin{abstract}
		\noindent
		We present the first direct excitation of the presumably ultra-narrow $1001 \, \textrm{nm}$ ground state transition in atomic dysproium. By using resonance ionization spectroscopy with pulsed Ti:sapphire lasers at a hot cavity laser ion source, we were able to measure the isotopic shifts in the $1001 \, \textrm{nm}$ line  between all seven stable isotopes. Furthermore, we determined the upper level energy from the atomic transition frequency of the $^{164} \textrm{Dy}$ isotope as $9991.004(1) \, \textrm{cm}^{-1}$ and confirm the level energy listed in the NIST database. Since a sufficiently narrow natural linewidth is an essential prerequisite for high precision spectroscopic investigations for fundamental questions we furthermore determined a lower limit of $2.9(1) \, \mu\textrm{s}$ for the lifetime of the excited state. 
		
	\end{abstract}
	
	\maketitle
	
	\section{Introduction}
	\noindent
	Narrow-linewidth atomic transitions can serve as highly sensitive probes for various inner and outer atomic interaction potentials and respective forces, and have become a general-purpose tool in the field of quantum many body physics \cite{Yamaguchi.2010, Martin.2013}. In addition, narrow linewidth is usually accompanied with long lifetimes of the excited states, such that these transitions offer precise, coherent control over metastable state populations and therefore allow for, e.g., the study of quantum gas mixtures and Kondo type physics \cite{FossFeig.2010, Riegger.2018} or the implementaition of qubits \cite{Daley.2008, Gorshkov.2009}. Beyond that, precision isotope shift measurements \cite{King.1963} have been suggested as vehicle to reveal high-energy physics contributions to atomic spectra and search for physics beyond the standard model \cite{Flambaum.2018, Berengut.2018, Delaunay.2017}.
	Various atomic species possess ultra-narrow transitions; however, dysprosium is a particularly interesting case. Due to its high magnetic moment and in consequence anisotropic long range interaction, dysprosium is highly attractive for quantum many body physics. On the other hand, many high-energy effects scale with atomic mass. Thus dysprosium with about 160 nucleons and 7 stable isotopes is an ideal study case.  Finding and characterizing a particularly narrow optical transitions in this system therefore is of high relevance. \\
	Some promising narrow-linewidth transitions are discussed in \cite{Dzuba.2010}, including calculated level energies and lifetimes which are compared to experimental values, if available. One is the  1001 nm $4\textrm{f}^{10}6\textrm{s}^2(^5\textrm{I}_8) \rightarrow 4\textrm{f}^9(^6\textrm{H}^\textrm{o})5\textrm{d}6\textrm{s}^2(^7\textrm{I}^\textrm{o}_9)$ ground state transition with a theoretically predicted linewidth of 53 Hz \cite{Dzuba.2010}. It was first observed indirectly in the spectrum of an induction lamp filled with $^{162}$Dy \cite{Conway.1971}. The NIST database \cite{Kramida.2018} reports an energy of  9990.97(1)\,\icm for the upper level, which corresponds to a transition wavelength of 1000.904(1)\,nm. In contrast calculations using the configuration interaction (CI) method yield a value of 9944\,\icm (corresponding to 1005.6\,nm) \cite{Dzuba.2010}. This and the fact that the transition at 1001\,nm could not be detected via fluorescence laser spectroscopy in an atomic beam \cite{Lu.2011b} motivate the verification of either result.\\
	In order to detect this weak transition we use the highly efficient technique of laser resonance ionization spectroscopy (RIS) \cite{Letokhov.1979}. The aim of this work is to (i) determine the exact transition frequency and extract first values for the isotope shifts of all stable isotopes, (ii) give a lower limit for the lifetime of the excited state, since a sufficiently narrow natural linewidth is a prerequisite for precision spectroscopy.\\
	
	\section{Experimental\label{sec:experimental_technique}}
	\subsection{Setup} \label{sec:setup}
\begin{figure}[t]
		\includegraphics[width=0.47\textwidth]{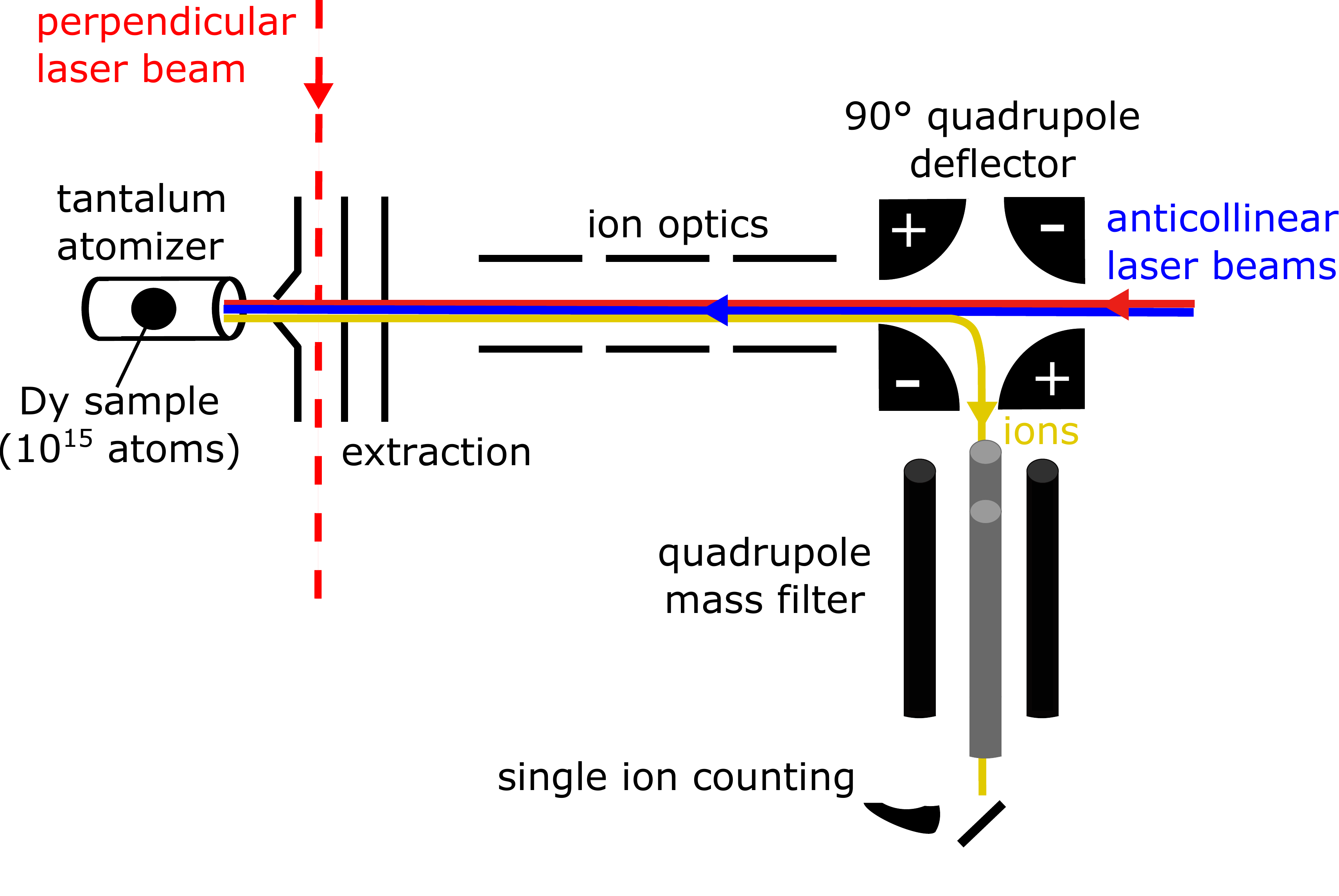}
		\caption{(color online) Sketch of the atomic beam mass spectrometer with ion flight path (yellow) and laser beams in anticollinear (solid red, solid blue) and perpendicular, crossed beam (dashed red) geometry. In the latter case the first extraction electrode is put on a positive voltage to act as an ion repeller.}
		\label{fig:mabu}
	\end{figure}

	\noindent
	Laser resonance ionization is based on multi-step photoionization via characteristic transitions of the element under investigation. Due to the typically high efficiency and selectivity, this technique is often applied at radioactive ion beam facilities, such as ISOLDE at CERN, both for ion beam production \cite{Fedosseev.2012, Lassen.2006} and spectroscopy of short-lived radioisotopes \cite{Cocolios.2013, Rothe.2013, Groote.2017}.\\
	Similarly our setup is optimized with respect to high sensitivity and relies on the hot-cavity laser ion source technique, combined with a low energy quadrupole mass spectrometer. Fig. \ref{fig:mabu} shows a sketch of the apparatus. A detailed description is given in \cite{Sonnenschein.2012}. 
	In our experiment we use a sample of $\approx 10^{15}$ Dy atoms, prepared from a standard nitric acid solution\footnote{Alfa Aesar Dysprosium AAS}, which is enclosed in a 3x3\,mm$^2$ Zr carrier foil and introduced into a tantalum oven with 35\,mm length and an inner diameter of 2\,mm. Dysprosium atoms are ionized by three  properly synchronized laser pulses at a repetition rate of 10\,kHz. The laser beams are overlapped anticollinearly with the ion beam axis, so that ionization takes place directly inside the atomizer oven. Alternatively, one may guide the laser beams through a side window of the vacuum chamber, perpendicularly intersecting the effusing atomic beam in front of the oven. This significantly reduces spectral Doppler broadening at the cost of approximately two orders of magnitude in ionization efficiency.\\
	The laser system consists of up to four pulsed Ti:sapphire lasers, each of them pumped with 12-18\,W average power of a commercial 532\,nm pulsed Nd:YAG laser\footnote{Photonics Industries DM100-532}. The Ti:sapphire lasers have pulse lengths of typically 50\,ns with up to 4\,W average output power. They can be tuned from about 680\,nm to 940\,nm and have a spectral linewidth of 1-10\,GHz depending on the specific resonator components used. The tuning range can be extended with second, third and fourth harmonic generation. Details of the laser system are given in \cite{Rothe.2011, Sonnenschein.2014, Wolf.2018}. For wide range scans we use a modified laser design, featuring a diffraction grating in Littrow configuration for frequency selection \cite{Teigelhofer.2010}. This laser type has an output power of up to 2 W and can be tuned mod-hop-free from 700\,nm to 1020\,nm. Under optimal conditions the linewidth is 1.5 GHz, however at wavelengths far from the Ti:sapphire gain maximum the pump power and the Ti:sapphire crystal position in the resonator have to be adjusted specifically so that the linewidth increases to 5\,GHz. The fundamental output of each laser is measured with a wavelength meter\footnote{High Finesse WS6-600 for wide range scans and High Finesse WSU-30 for isotope shift measurements}.\\

	\subsection{Spectroscopic Technique \label{sec:spectroscopy_technique}}
\begin{figure}[b]
		\includegraphics[width=0.495\textwidth]{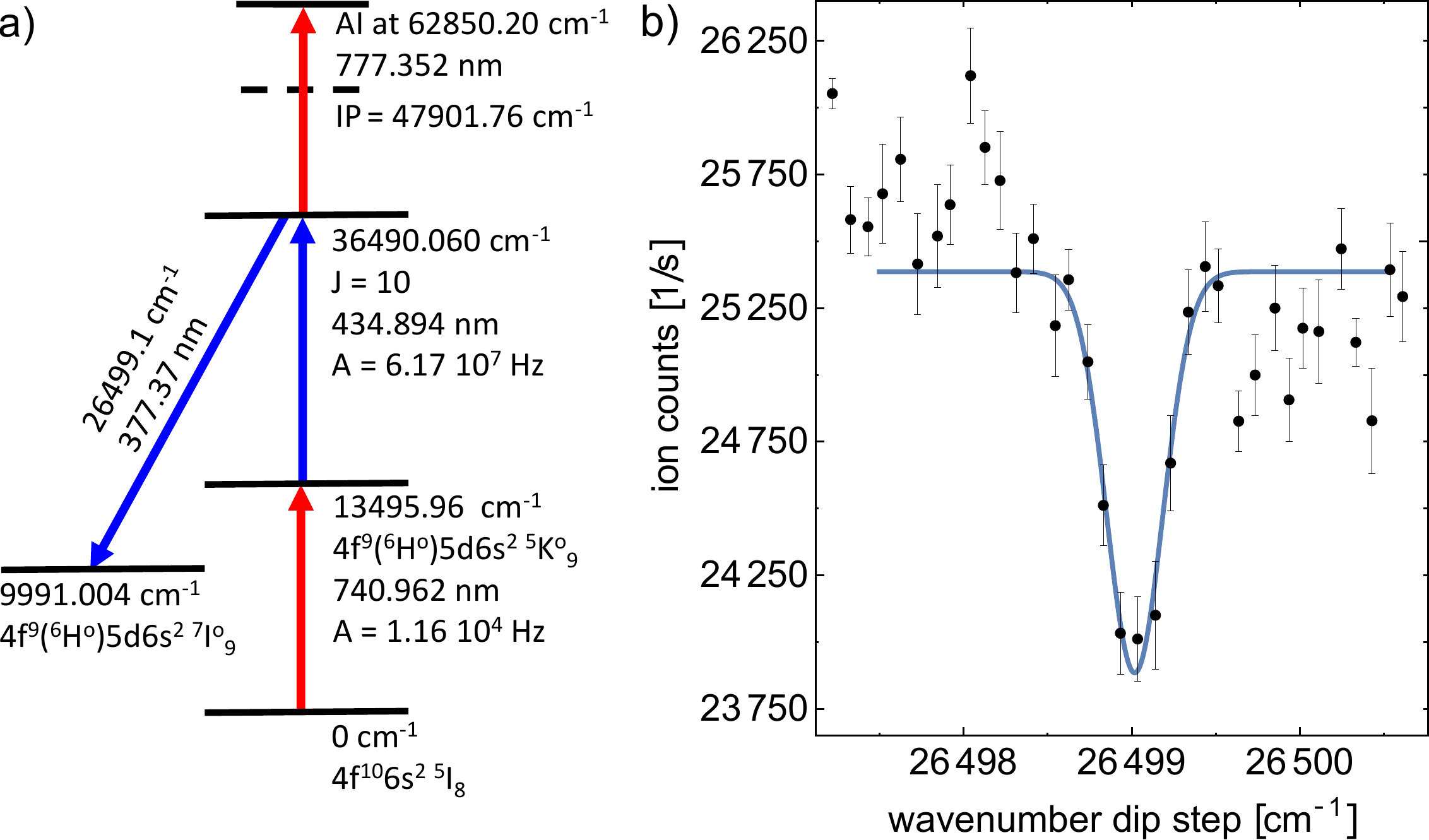}
		\caption{(color online) Indirect measurement of the upper energy level belonging to the $1001 \, \textrm{nm}$ transition. a) Measurement scheme consisting of a three step resonant excitation ending above the first ionization potential $\textrm{IP}=47901.76 \, \textrm{cm}^{-1}$ of dysprosium \cite{Studer.2017} together with the additional dip step.  b) Ion counts as a function of the dip step wavenumber.}
		\label{fig:Dip_measurement}
	\end{figure}

	\noindent
	Spectroscopy is performed by detuning one laser excitation step while monitoring the ion count rate. Since this requires a photoionization scheme to begin with, the initial measurements were carried out with a scheme based on the $4\textrm{f}^{10}6\textrm{s}^2(^5\textrm{I}_8) \rightarrow 4\textrm{f}^9(^6\textrm{H}^\textrm{o})5\textrm{d}6\textrm{s}^2(^5\textrm{K}^\textrm{o}_9)$ ground state transition at 741\,nm. The excited state has a similar configuration to the one at 1001\,nm, but the line intensity is a factor of $\approx$ 50 higher \cite{Lu.2011, Dzuba.2010}. This enormously facilitates the development of a full resonant three-step ionization scheme. Starting from the excited state of the 741\,nm transition, a wide range spectrum of high-lying states was accessed by scanning the second harmonics of a grating-assisted Ti:sapphire laser between 401\,nm and 437\,nm. A third laser at 780\,nm (the gain maximum of Ti:sapphire) is used for non-resonant ionization of excited atoms. The spectrum shows over 100 lines, with some of the upper states known in literature \cite{Kramida.2018}. A complete list of recorded lines is given in the supplementary material \cite{Supplement}. In most cases a resonant third excitation step to an auto-ionizing state can be easily found in the dense spectrum of dysprosium by detuning the ionization laser output by few nanometers.\\
	To connect our photoionization scheme to the 1001\,nm transition, we use a fourth laser to de-excitate atoms from the second excited state into the $4\textrm{f}^9(^6\textrm{H}^\textrm{o})5\textrm{d}6\textrm{s}^2(^7\textrm{I}^\textrm{o}_9)$ state. When the laser is resonant to the transition, the de-excitation competes with the ionization and a dip in the ion signal can be observed, as shown in FIG. \ref{fig:Dip_measurement}. \\
	The rather weak dip signal of less than 10\,\% of the total ion counts may be related to the fact that the specifically induced deexcitation competes with a number of loss channels through spontaneous decay into other lower lying levels. With the dip-technique we can not only indirectly measure the level energy of the excited state, but also prepare for a full-resonant three-step ionization scheme for the direct excitation of the 1001\,nm ground state transition. To further reduce the uncertainty of the $4\textrm{f}^9(^6\textrm{H}^\textrm{o})5\textrm{d}6\textrm{s}^2(^7\textrm{I}^\textrm{o}_9)$ level energy, which depends on three wavelengths in the dip measurements, we proceed by using the settled ionization scheme involving the direct excitation of the 1001\,nm transition.
	
	\section{Direct excitation spectroscopy}
	
	\subsection{1001\,nm transition} \label{sec:spectroscopy_1000}
\begin{figure}[b]
		\begin{center}
			\includegraphics[width=0.495\textwidth]{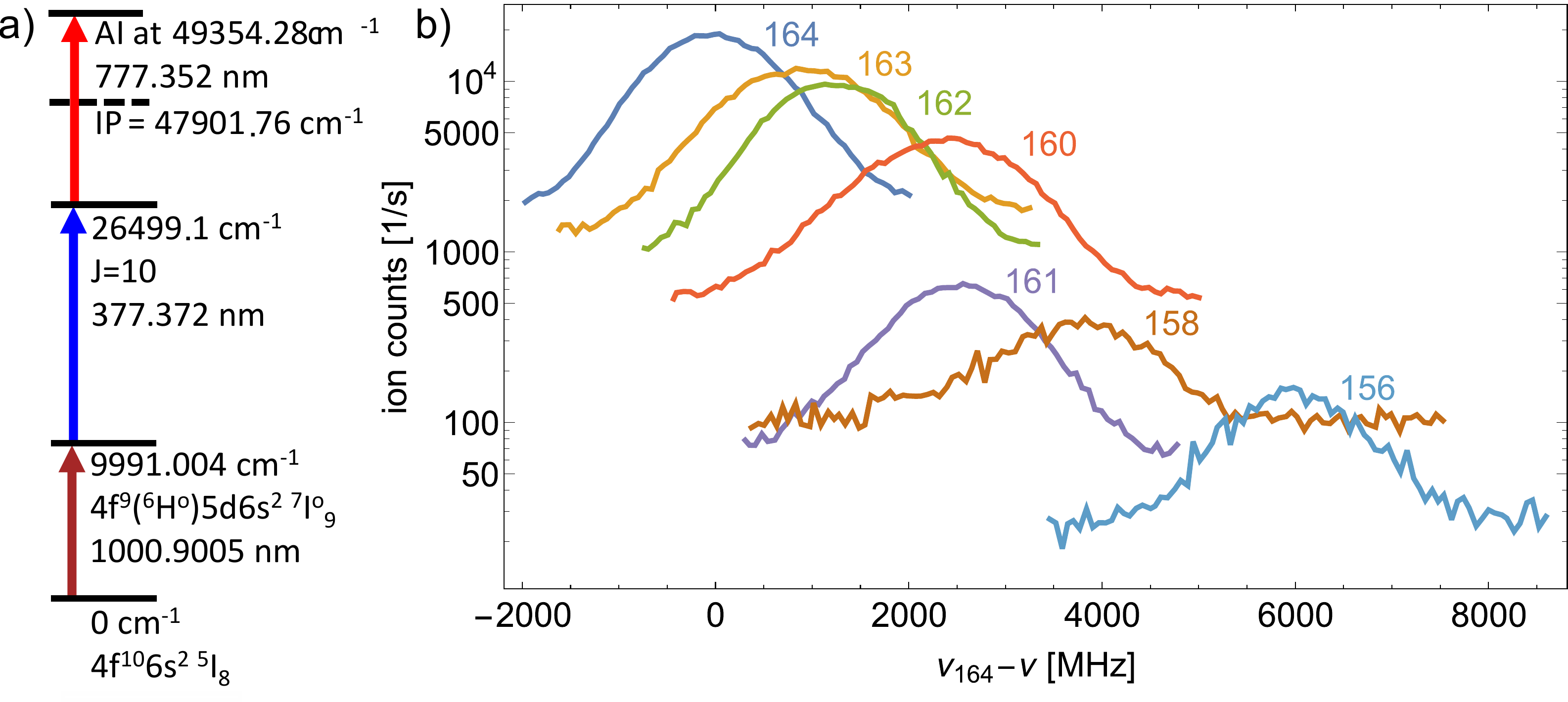}
			\caption{(color online) Isotope shift measurement for the $1001 \, \textrm{nm}$ ground state transition. a) Excitation scheme. b) Ion counts as a function of first excitation step frequency $\nu$ relative to the atomic transition frequency  ${\nu_{164} = 299.5228(1) \, \textrm{THz}}$ of the $^{164}\textrm{Dy}$ isotope.}
			\label{fig:isotope_shift_1001}
		\end{center}
	\end{figure}

	\noindent
	In this section we discuss the first direct excitation of the $1001 \, \textrm{nm}$ ground state transition. For the measurement all laser beams are oriented anticollinearly to the atom beam (see FIG. \ref{fig:mabu}). A perpendicular geometry does not lead to an improvement at this point, since the expected doppler broadening of $\approx 600 \, \textrm{MHz}$ within the tantalum oven is in the order of the laser linewidth. The beam diameter at atom position is about $2 \, \textrm{mm}$ which corresponds to the inner diameter of the oven. For the photoionization a Ti:sapphire laser with a wavelength of $1001 \,\textrm{nm}$ first excites the atoms into the $4\textrm{f}^9(^6\textrm{H}^\textrm{o})5\textrm{d}6\textrm{s}^2(^7\textrm{I}^\textrm{o}_9)$ state. From this state the atoms are resonantly excited to the second excited state with an energy of $26499.1 \, \textrm{cm}^{-1}$ with a wavelength of $377 \, \textrm{nm}$ by using a second, this time frequency doubled  Ti:sapphire laser. In the last step a third laser with a wavelength of $777 \, \textrm{nm}$ addresses an autoionizing state to resonantly ionize the atoms. 
	The complete excitation scheme is shown in FIG. \ref{fig:isotope_shift_1001} a). \\
	While the frequencies of the upper two steps are fixed for the whole measurement, the frequency of the first step is scanned. Due to its wide tuning range the grating assisted laser, as described in SEC. \ref{sec:setup}, was used to probe the 1001\,nm transition, while an additional etalon ($d=2$\,mm, $R=0.4$) was inserted to reduce the linewidth to 1\,GHz. FIG. \ref{fig:isotope_shift_1001} b) shows the resulting direct excitation of all seven stable dysprosium isotopes. For the individual isotopes we adapted the mass spectrometer setting accordingly. \\ From the measurement we calculated the isotopic shifts in the $1001 \, \textrm{nm}$ transition which are listed in TABLE \ref{tab:isotope_shift_1001}. 
	\begin{table}[h]
		\renewcommand{\arraystretch}{1.3}
		\caption{Isotope shift in the $1001 \, \textrm{nm}$ transition
			in Dy, relative to the isotope 164.}
		\begin{tabular}{c|c} \hline \hline
			& isotope shift $\delta \nu$ [MHz]\\ \hline 
			$\delta \nu_{164-163}$ & 907(36)\\ 
			$\delta \nu_{164-162}$ & 1233(35)\\ 
			$\delta \nu_{164-161}$ & 2337(37)\\ 
			$\delta \nu_{164-160}$ & 2566(36)\\ 
			$\delta \nu_{164-158}$ & 3685(45)\\
			$\delta \nu_{164-156}$ & 5976(39)\\ \hline \hline 
		\end{tabular}
		\label{tab:isotope_shift_1001}
	\end{table}
	The specified error corresponds to the sum of the fit error and an estimated error of $30 \, \textrm{MHz}$ for the drift of the wavelength meter during the measuring time. Any other systematics are comparatively small and were neglected. The error estimation is based on later measurements in which we determined the drift as $\approx 10 \, \textrm{MHz}$ per hour as well as the long measuring time due to the individual mass spectrometer setting for each isotope. Furthermore, we determined the upper level energy from the atomic transition frequency of the $^{164} \textrm{Dy}$ isotope as $9991.004(1) \, \textrm{cm}^{-1}$.  Taking the isotope shift into account the latter is in good agreement with the value listed in the NIST database \cite{Kramida.2018}.
	
	\subsection{741 nm transition\label{sec:spectroscopy_741}} 
\begin{figure}[t]
		\includegraphics[width=0.495\textwidth]{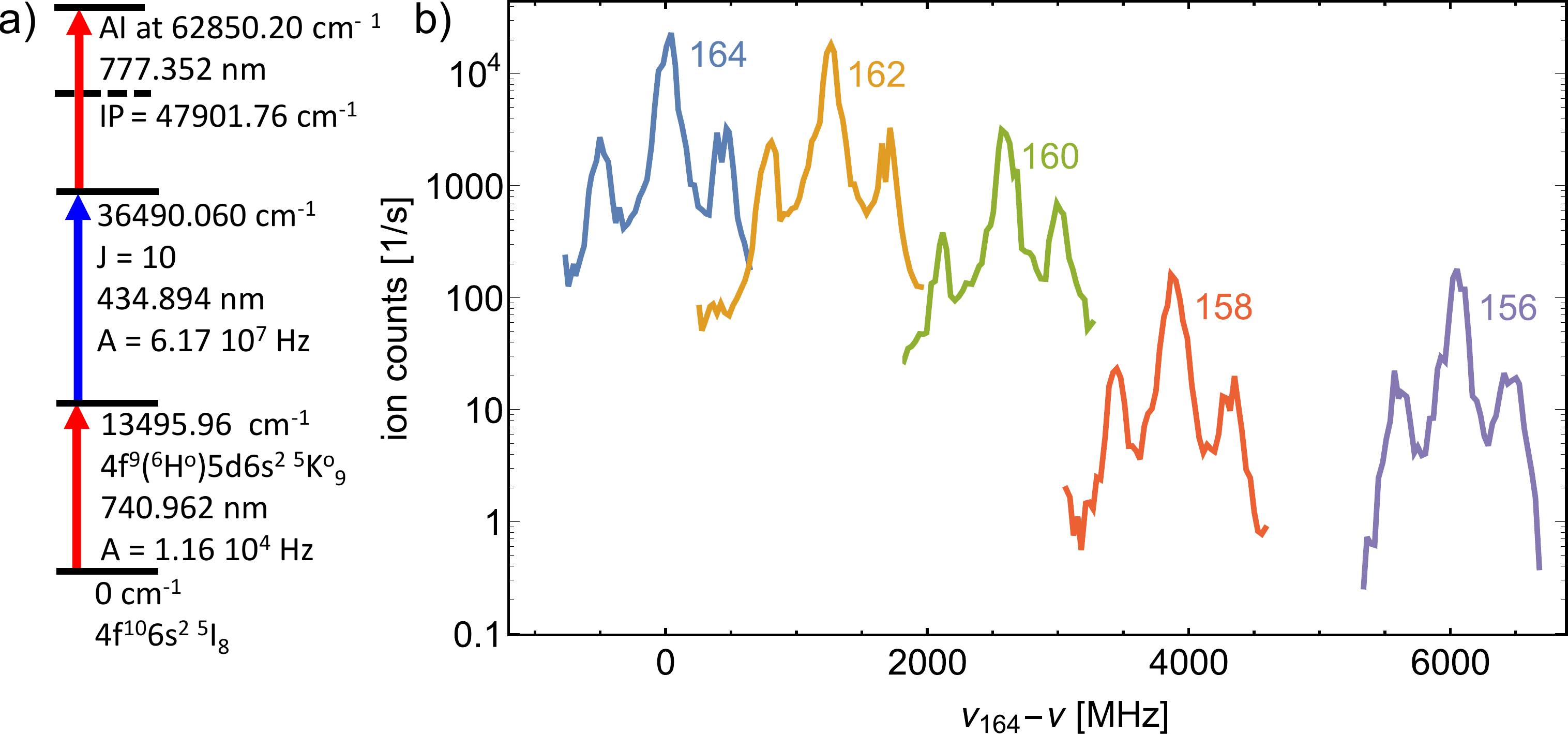}
		\caption{(color online) Isotope shift measurement for the $741 \, \textrm{nm}$ ground state transition.  a) Excitation scheme. b) Ion counts as a function of first excitation step frequency $\nu$ relative to the atomic transition frequency  ${\nu_{164}= 404.5990(1) \, \textrm{THz}}$ of the $^{164}\textrm{Dy}$ isotope. Laser side modes are clearly visible.}
		\label{fig:isotope_shift_741}
	\end{figure}

	\noindent
	In the course of the indirect detection of the upper energy level belonging to the $1001 \, \textrm{nm}$ transition we were also able to measure the isotope shift in the first excitation step  along the $741 \, \textrm{nm}$ ground state transition. These data were measured with a different implementation of the standard Ti:sapphire laser featuring a bowtie resonator design. Similar to the standard laser, frequency selection is achieved by a combination of a birefringent filter and a solid etalon ($d=0.3$ mm, $R=0.4$), but with the option to add an additional piezo-actuated air-spaced etalon ($d=12$ mm, $R=0.4$). This potentially allows operation on a single longitudinal mode, however since the cavity lacks active stabilization at this stage it suffers from an occasional rise of side-modes, which appear at $\delta \nu= \pm 443 \, \textrm{MHz}$ and are suppressed by only about a factor of $\approx 10$.\\
	FIG. \ref{fig:isotope_shift_741} shows the direct excitation of the five stable bosonic isotopes and TABLE \ref{tab:isotope_shift_741} the resulting isotope shifts. Since the isotope shifts of the stable odd-mass isotopes with non-zero nuclear spin $I$ as well as the isotope shifts of the three stable even-mass isotopes (${I = 0}$) with highest abundance are already given in \cite{Lu.2011}, we omitted a re-measurement of the odd-mass isotopes at this point. The isotope shifts obtained here for the even-mass isotopes with highest abundance are in accordance with \cite{Lu.2011}. For the two rarest stable isotopes $^{158}\textrm{Dy}$ and $^{156}\textrm{Dy}$ this was the first isotope shift measurement in this line. \\
	
	\begin{table}[h]
		\renewcommand{\arraystretch}{1.4}
		\caption{Isotope shift in the $741 \, \textrm{nm}$ transition
			in Dy for the five isotopes with even mass number.}
		\begin{tabular}{c|c} \hline \hline
			& isotope shift $\delta \nu$ [MHz]\\ \hline 
			$\delta \nu_{164-162}$ & 1245(32)\\ 
			$\delta \nu_{164-160}$ & 2583(32)\\ 
			$\delta \nu_{164-158}$ & 3874(32)\\
			$\delta \nu_{164-156}$ & 6042(32)\\ \hline \hline 
		\end{tabular}
		\label{tab:isotope_shift_741}
	\end{table}

	\section{Lifetime measurements}
\begin{center}
		\begin{figure}[t]
			\includegraphics[width=0.38\textwidth]{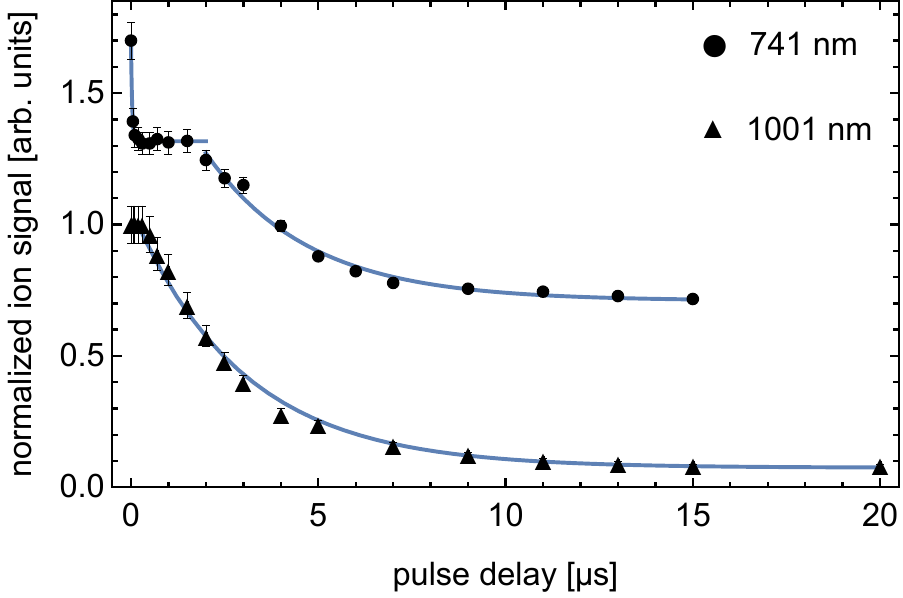}
			\caption{(color online) Lifetime measurements in the 741 nm and 1001 nm transitions obtained by delaying the ionization laser pulse. For better readability we added a random offset to the ion signal of the $741 \, \textrm{nm}$ transition.}
			\label{fig:lifetime}
		\end{figure}
	\end{center}

	\noindent
	Lower limits for the lifetimes of the excited states at 9990.96\,\icm and 13495.96\,\icm can be determined by investigating the population decay after the pulsed laser excitation. The population is probed by delayed ionization laser pulses, according to the excitation schemes in section \ref{sec:spectroscopy_1000} and \ref{sec:spectroscopy_741}, and subsequent ion counting. The temporal profiles and delays between the individual Ti:sapphire laser pulses are captured by fast photodiodes and monitored with an oscilloscope. In order to maintain stable laser power during the measurement, the first excitation step laser is pumped by a separate Nd:YAG laser\footnote{Quantronix Hawk-Pro 532-60-M} and pulse delays are controlled by means of shifting the pump laser trigger  accordingly. FIG. \ref{fig:lifetime} shows the excited state population decay at oven temperatures of 700(100)\,\celsius and 950(100)\,\celsius for the states at 9990.96\,\icm and 13495.96\,\icm, respectively. The excited state at 9990.96 \icm shows an exponential decay with a lifetime of 2.9(1)\,$\mu$s, whereas the 13495.96\,\icm state clearly features two components with lifetimes of 32(2)\,ns and 2.8(3)\,$\mu$s. The short-lived contribution is an artifact, which is related to the ionization scheme for the 741\,nm transition where the first excitation step has enough energy to ionize atoms parasitically from the second excited state. This effect is completely suppressed by a delayed second excitation pulse, thus the 32(2)\,ns lifetime corresponds to the laser pulse length. Nonetheless, lifetimes of the investigated states are orders of magnitude shorter than theoretical values \cite{Dzuba.2010}, which actually was expected due to the experimental circumstances. De-excitation of atoms within the hot atomic vapor may occur by collisions with the oven wall or other atoms. In comparison to an experimental value for the lifetime of the 13495.96 \icm excited state of 89.3(8)\,$\mu$s \cite{Lu.2011} and the fact that both of our measured values agree with each other we can conclude that, in our experiment, the extracted lifetimes predominately depend on the mean free path of the hot atoms within the atomic beam oven cavity. Consequently the 2.9(1)\,$\mu$s lifetime for the 9990.96\,\icm excited state should be treated as a very conservative lower limit, which corresponds to an upper limit for the 1001\,nm transition linewidth of 55(2)\,kHz.

	\section{Conclusion}
	\noindent
	We presented the first direct excitation of the  $1001 \, \textrm{nm}$ ground state transition for all seven stable dysprosium isotopes, applying high repetition rate pulsed laser resonance ionization. Furthermore we measured the isotopic shift in the $1001 \, \textrm{nm}$ transition and determined the upper level energy from the atomic transition frequency of the $^{164} \textrm{Dy}$ isotope as $9991.004(1) \, \textrm{cm}^{-1}$, which is in accordance with the value listed in the NIST database \cite{Kramida.2018}. We obtain an upper limit of $55(2) \, \textrm{kHz}$ for the transition linewidth of the excited state.  Our results open the route towards investigation of physics beyond the standard model of particle physics and enable the study of many body physics with magnetic quantum gases by atomic high resolution spectroscopy within the $1001 \, \textrm{nm}$ transition.

	\section{Acknowledgment}
	\noindent
	The authors would like to point out that this work is part of the M.Sc. thesis of L. Maske, who not only contributed to the experimental activities, but also did a major part of the data evaluation. We thank N. Petersen and F. M\"uhlbauer for their ideas and fruitful discussions. We gratefully acknowledge the financial support of the EU through ENSAR2-RESIST (Grant no. 654002) and DFG Gro\ss{}ger\"at: DFG FUGG (INST 247/818-1).

\bibliography{Dy_1000}

\begin{thebibliography}{28}%
\makeatletter
\providecommand \@ifxundefined [1]{%
 \@ifx{#1\undefined}
}%
\providecommand \@ifnum [1]{%
 \ifnum #1\expandafter \@firstoftwo
 \else \expandafter \@secondoftwo
 \fi
}%
\providecommand \@ifx [1]{%
 \ifx #1\expandafter \@firstoftwo
 \else \expandafter \@secondoftwo
 \fi
}%
\providecommand \natexlab [1]{#1}%
\providecommand \enquote  [1]{``#1''}%
\providecommand \bibnamefont  [1]{#1}%
\providecommand \bibfnamefont [1]{#1}%
\providecommand \citenamefont [1]{#1}%
\providecommand \href@noop [0]{\@secondoftwo}%
\providecommand \href [0]{\begingroup \@sanitize@url \@href}%
\providecommand \@href[1]{\@@startlink{#1}\@@href}%
\providecommand \@@href[1]{\endgroup#1\@@endlink}%
\providecommand \@sanitize@url [0]{\catcode `\\12\catcode `\$12\catcode
  `\&12\catcode `\#12\catcode `\^12\catcode `\_12\catcode `\%12\relax}%
\providecommand \@@startlink[1]{}%
\providecommand \@@endlink[0]{}%
\providecommand \url  [0]{\begingroup\@sanitize@url \@url }%
\providecommand \@url [1]{\endgroup\@href {#1}{\urlprefix }}%
\providecommand \urlprefix  [0]{URL }%
\providecommand \Eprint [0]{\href }%
\providecommand \doibase [0]{http://dx.doi.org/}%
\providecommand \selectlanguage [0]{\@gobble}%
\providecommand \bibinfo  [0]{\@secondoftwo}%
\providecommand \bibfield  [0]{\@secondoftwo}%
\providecommand \translation [1]{[#1]}%
\providecommand \BibitemOpen [0]{}%
\providecommand \bibitemStop [0]{}%
\providecommand \bibitemNoStop [0]{.\EOS\space}%
\providecommand \EOS [0]{\spacefactor3000\relax}%
\providecommand \BibitemShut  [1]{\csname bibitem#1\endcsname}%
\let\auto@bib@innerbib\@empty
\bibitem [{\citenamefont {Yamaguchi}\ \emph {et~al.}(2010)\citenamefont
  {Yamaguchi}, \citenamefont {Uetake}, \citenamefont {Kato}, \citenamefont
  {Ito},\ and\ \citenamefont {Takahashi}}]{Yamaguchi.2010}%
  \BibitemOpen
  \bibfield  {author} {\bibinfo {author} {\bibfnamefont {A.}~\bibnamefont
  {Yamaguchi}}, \bibinfo {author} {\bibfnamefont {S.}~\bibnamefont {Uetake}},
  \bibinfo {author} {\bibfnamefont {S.}~\bibnamefont {Kato}}, \bibinfo {author}
  {\bibfnamefont {H.}~\bibnamefont {Ito}}, \ and\ \bibinfo {author}
  {\bibfnamefont {Y.}~\bibnamefont {Takahashi}},\ }\href {\doibase
  \url{10.1088/1367-2630/12/10/103001}} {\bibfield  {journal} {\bibinfo
  {journal} {{New J. Phys. (New Journal of Physica)}}\ }\textbf {\bibinfo
  {volume} {12}},\ \bibinfo {pages} {103001} (\bibinfo {year}
  {2010})}\BibitemShut {NoStop}%
\bibitem [{\citenamefont {Martin}\ \emph {et~al.}(2013)\citenamefont {Martin},
  \citenamefont {Bishof}, \citenamefont {Swallows}, \citenamefont {Zhang},
  \citenamefont {Benko}, \citenamefont {von Stecher}, \citenamefont {Gorshkov},
  \citenamefont {Rey},\ and\ \citenamefont {Ye}}]{Martin.2013}%
  \BibitemOpen
  \bibfield  {author} {\bibinfo {author} {\bibfnamefont {M.~J.}\ \bibnamefont
  {Martin}}, \bibinfo {author} {\bibfnamefont {M.}~\bibnamefont {Bishof}},
  \bibinfo {author} {\bibfnamefont {M.~D.}\ \bibnamefont {Swallows}}, \bibinfo
  {author} {\bibfnamefont {X.}~\bibnamefont {Zhang}}, \bibinfo {author}
  {\bibfnamefont {C.}~\bibnamefont {Benko}}, \bibinfo {author} {\bibfnamefont
  {J.}~\bibnamefont {von Stecher}}, \bibinfo {author} {\bibfnamefont {A.~V.}\
  \bibnamefont {Gorshkov}}, \bibinfo {author} {\bibfnamefont {A.~M.}\
  \bibnamefont {Rey}}, \ and\ \bibinfo {author} {\bibfnamefont
  {J.}~\bibnamefont {Ye}},\ }\href {\doibase \url{10.1126/science.1236929}}
  {\bibfield  {journal} {\bibinfo  {journal} {{Science (N.Y.)}}\ }\textbf
  {\bibinfo {volume} {341}},\ \bibinfo {pages} {632} (\bibinfo {year}
  {2013})}\BibitemShut {NoStop}%
\bibitem [{\citenamefont {Foss-Feig}\ \emph {et~al.}(2010)\citenamefont
  {Foss-Feig}, \citenamefont {Hermele},\ and\ \citenamefont
  {Rey}}]{FossFeig.2010}%
  \BibitemOpen
  \bibfield  {author} {\bibinfo {author} {\bibfnamefont {M.}~\bibnamefont
  {Foss-Feig}}, \bibinfo {author} {\bibfnamefont {M.}~\bibnamefont {Hermele}},
  \ and\ \bibinfo {author} {\bibfnamefont {A.~M.}\ \bibnamefont {Rey}},\ }\href
  {\doibase \url{10.1103/PhysRevA.81.051603}} {\bibfield  {journal} {\bibinfo
  {journal} {{Physical Review A}}\ }\textbf {\bibinfo {volume} {81}},\ \bibinfo
  {pages} {051603(R)} (\bibinfo {year} {2010})}\BibitemShut {NoStop}%
\bibitem [{\citenamefont {Riegger}\ \emph {et~al.}(2018)\citenamefont
  {Riegger}, \citenamefont {{Darkwah Oppong}}, \citenamefont {H{\"o}fer},
  \citenamefont {Fernandes}, \citenamefont {Bloch},\ and\ \citenamefont
  {F{\"o}lling}}]{Riegger.2018}%
  \BibitemOpen
  \bibfield  {author} {\bibinfo {author} {\bibfnamefont {L.}~\bibnamefont
  {Riegger}}, \bibinfo {author} {\bibfnamefont {N.}~\bibnamefont {{Darkwah
  Oppong}}}, \bibinfo {author} {\bibfnamefont {M.}~\bibnamefont {H{\"o}fer}},
  \bibinfo {author} {\bibfnamefont {D.~R.}\ \bibnamefont {Fernandes}}, \bibinfo
  {author} {\bibfnamefont {I.}~\bibnamefont {Bloch}}, \ and\ \bibinfo {author}
  {\bibfnamefont {S.}~\bibnamefont {F{\"o}lling}},\ }\href {\doibase
  \url{10.1103/PhysRevLett.120.143601}} {\bibfield  {journal} {\bibinfo
  {journal} {{Physical Review Letters}}\ }\textbf {\bibinfo {volume} {120}},\
  \bibinfo {pages} {143601} (\bibinfo {year} {2018})}\BibitemShut {NoStop}%
\bibitem [{\citenamefont {Daley}\ \emph {et~al.}(2008)\citenamefont {Daley},
  \citenamefont {Boyd}, \citenamefont {Ye},\ and\ \citenamefont
  {Zoller}}]{Daley.2008}%
  \BibitemOpen
  \bibfield  {author} {\bibinfo {author} {\bibfnamefont {A.~J.}\ \bibnamefont
  {Daley}}, \bibinfo {author} {\bibfnamefont {M.~M.}\ \bibnamefont {Boyd}},
  \bibinfo {author} {\bibfnamefont {J.}~\bibnamefont {Ye}}, \ and\ \bibinfo
  {author} {\bibfnamefont {P.}~\bibnamefont {Zoller}},\ }\href {\doibase
  \url{10.1103/PhysRevLett.101.170504}} {\bibfield  {journal} {\bibinfo
  {journal} {{Physical Review Letters}}\ }\textbf {\bibinfo {volume} {101}},\
  \bibinfo {pages} {170504} (\bibinfo {year} {2008})}\BibitemShut {NoStop}%
\bibitem [{\citenamefont {Gorshkov}\ \emph {et~al.}(2009)\citenamefont
  {Gorshkov}, \citenamefont {Rey}, \citenamefont {Daley}, \citenamefont {Boyd},
  \citenamefont {Ye}, \citenamefont {Zoller},\ and\ \citenamefont
  {Lukin}}]{Gorshkov.2009}%
  \BibitemOpen
  \bibfield  {author} {\bibinfo {author} {\bibfnamefont {A.~V.}\ \bibnamefont
  {Gorshkov}}, \bibinfo {author} {\bibfnamefont {A.~M.}\ \bibnamefont {Rey}},
  \bibinfo {author} {\bibfnamefont {A.~J.}\ \bibnamefont {Daley}}, \bibinfo
  {author} {\bibfnamefont {M.~M.}\ \bibnamefont {Boyd}}, \bibinfo {author}
  {\bibfnamefont {J.}~\bibnamefont {Ye}}, \bibinfo {author} {\bibfnamefont
  {P.}~\bibnamefont {Zoller}}, \ and\ \bibinfo {author} {\bibfnamefont {M.~D.}\
  \bibnamefont {Lukin}},\ }\href {\doibase
  \url{10.1103/PhysRevLett.102.110503}} {\bibfield  {journal} {\bibinfo
  {journal} {{Physical Review Letters}}\ }\textbf {\bibinfo {volume} {102}},\
  \bibinfo {pages} {110503} (\bibinfo {year} {2009})}\BibitemShut {NoStop}%
\bibitem [{\citenamefont {King}(1963)}]{King.1963}%
  \BibitemOpen
  \bibfield  {author} {\bibinfo {author} {\bibfnamefont {W.~H.}\ \bibnamefont
  {King}},\ }\href {\doibase \url{10.1364/JOSA.53.000638}} {\bibfield
  {journal} {\bibinfo  {journal} {{Journal of the Optical Society of America}}\
  }\textbf {\bibinfo {volume} {53}},\ \bibinfo {pages} {638} (\bibinfo {year}
  {1963})}\BibitemShut {NoStop}%
\bibitem [{\citenamefont {Flambaum}\ \emph {et~al.}(2018)\citenamefont
  {Flambaum}, \citenamefont {Geddes},\ and\ \citenamefont
  {Viatkina}}]{Flambaum.2018}%
  \BibitemOpen
  \bibfield  {author} {\bibinfo {author} {\bibfnamefont {V.~V.}\ \bibnamefont
  {Flambaum}}, \bibinfo {author} {\bibfnamefont {A.~J.}\ \bibnamefont
  {Geddes}}, \ and\ \bibinfo {author} {\bibfnamefont {A.~V.}\ \bibnamefont
  {Viatkina}},\ }\href {\doibase \url{10.1103/PhysRevA.97.032510}} {\bibfield
  {journal} {\bibinfo  {journal} {{Physical Review A}}\ }\textbf {\bibinfo
  {volume} {97}},\ \bibinfo {pages} {032510} (\bibinfo {year}
  {2018})}\BibitemShut {NoStop}%
\bibitem [{\citenamefont {Berengut}\ \emph {et~al.}(2018)\citenamefont
  {Berengut}, \citenamefont {Budker}, \citenamefont {Delaunay}, \citenamefont
  {Flambaum}, \citenamefont {Frugiuele}, \citenamefont {Fuchs}, \citenamefont
  {Grojean}, \citenamefont {Harnik}, \citenamefont {Ozeri}, \citenamefont
  {Perez},\ and\ \citenamefont {Soreq}}]{Berengut.2018}%
  \BibitemOpen
  \bibfield  {author} {\bibinfo {author} {\bibfnamefont {J.~C.}\ \bibnamefont
  {Berengut}}, \bibinfo {author} {\bibfnamefont {D.}~\bibnamefont {Budker}},
  \bibinfo {author} {\bibfnamefont {C.}~\bibnamefont {Delaunay}}, \bibinfo
  {author} {\bibfnamefont {V.~V.}\ \bibnamefont {Flambaum}}, \bibinfo {author}
  {\bibfnamefont {C.}~\bibnamefont {Frugiuele}}, \bibinfo {author}
  {\bibfnamefont {E.}~\bibnamefont {Fuchs}}, \bibinfo {author} {\bibfnamefont
  {C.}~\bibnamefont {Grojean}}, \bibinfo {author} {\bibfnamefont
  {R.}~\bibnamefont {Harnik}}, \bibinfo {author} {\bibfnamefont
  {R.}~\bibnamefont {Ozeri}}, \bibinfo {author} {\bibfnamefont
  {G.}~\bibnamefont {Perez}}, \ and\ \bibinfo {author} {\bibfnamefont
  {Y.}~\bibnamefont {Soreq}},\ }\href {\doibase
  \url{10.1103/PhysRevLett.120.091801}} {\bibfield  {journal} {\bibinfo
  {journal} {{Physical Review Letters}}\ }\textbf {\bibinfo {volume} {120}},\
  \bibinfo {pages} {091801} (\bibinfo {year} {2018})}\BibitemShut {NoStop}%
\bibitem [{\citenamefont {Delaunay}\ \emph {et~al.}(2017)\citenamefont
  {Delaunay}, \citenamefont {Ozeri}, \citenamefont {Perez},\ and\ \citenamefont
  {Soreq}}]{Delaunay.2017}%
  \BibitemOpen
  \bibfield  {author} {\bibinfo {author} {\bibfnamefont {C.}~\bibnamefont
  {Delaunay}}, \bibinfo {author} {\bibfnamefont {R.}~\bibnamefont {Ozeri}},
  \bibinfo {author} {\bibfnamefont {G.}~\bibnamefont {Perez}}, \ and\ \bibinfo
  {author} {\bibfnamefont {Y.}~\bibnamefont {Soreq}},\ }\href {\doibase
  \url{10.1103/PhysRevD.96.093001}} {\bibfield  {journal} {\bibinfo  {journal}
  {{Physical Review D}}\ }\textbf {\bibinfo {volume} {96}},\ \bibinfo {pages}
  {093001} (\bibinfo {year} {2017})}\BibitemShut {NoStop}%
\bibitem [{\citenamefont {Dzuba}\ and\ \citenamefont
  {Flambaum}(2010)}]{Dzuba.2010}%
  \BibitemOpen
  \bibfield  {author} {\bibinfo {author} {\bibfnamefont {V.~A.}\ \bibnamefont
  {Dzuba}}\ and\ \bibinfo {author} {\bibfnamefont {V.~V.}\ \bibnamefont
  {Flambaum}},\ }\href {\doibase \url{10.1103/PhysRevA.81.052515}} {\bibfield
  {journal} {\bibinfo  {journal} {{Physical Review A}}\ }\textbf {\bibinfo
  {volume} {81}},\ \bibinfo {pages} {052515} (\bibinfo {year}
  {2010})}\BibitemShut {NoStop}%
\bibitem [{\citenamefont {Conway}\ and\ \citenamefont
  {Worden}(1971)}]{Conway.1971}%
  \BibitemOpen
  \bibfield  {author} {\bibinfo {author} {\bibfnamefont {J.~G.}\ \bibnamefont
  {Conway}}\ and\ \bibinfo {author} {\bibfnamefont {E.~F.}\ \bibnamefont
  {Worden}},\ }\href {\doibase \url{10.1364/JOSA.61.000704}} {\bibfield
  {journal} {\bibinfo  {journal} {{Journal of the Optical Society of America}}\
  }\textbf {\bibinfo {volume} {61}},\ \bibinfo {pages} {704} (\bibinfo {year}
  {1971})}\BibitemShut {NoStop}%
\bibitem [{\citenamefont {Kramida}\ \emph {et~al.}(2018)\citenamefont
  {Kramida}, \citenamefont {Ralchenko}, \citenamefont {Reader},\ and\
  \citenamefont {{the NIST ASD Team}}}]{Kramida.2018}%
  \BibitemOpen
  \bibfield  {author} {\bibinfo {author} {\bibfnamefont {A.}~\bibnamefont
  {Kramida}}, \bibinfo {author} {\bibfnamefont {Y.}~\bibnamefont {Ralchenko}},
  \bibinfo {author} {\bibfnamefont {J.}~\bibnamefont {Reader}}, \ and\ \bibinfo
  {author} {\bibnamefont {{the NIST ASD Team}}},\ }\href
  {\url{http://physics.nist.gov/asd}} {\enquote {\bibinfo {title} {{NIST Atomic
  Spectra Database}},}\ } (\bibinfo {year} {2018})\BibitemShut {NoStop}%
\bibitem [{\citenamefont {Lu}\ \emph {et~al.}(2011{\natexlab{a}})\citenamefont
  {Lu}, \citenamefont {Burdick}, \citenamefont {Youn},\ and\ \citenamefont
  {Lev}}]{Lu.2011b}%
  \BibitemOpen
  \bibfield  {author} {\bibinfo {author} {\bibfnamefont {M.}~\bibnamefont
  {Lu}}, \bibinfo {author} {\bibfnamefont {N.~Q.}\ \bibnamefont {Burdick}},
  \bibinfo {author} {\bibfnamefont {S.~H.}\ \bibnamefont {Youn}}, \ and\
  \bibinfo {author} {\bibfnamefont {B.~L.}\ \bibnamefont {Lev}},\ }\href
  {\doibase \url{10.1103/PhysRevLett.107.190401}} {\bibfield  {journal}
  {\bibinfo  {journal} {{Physical Review Letters}}\ }\textbf {\bibinfo {volume}
  {107}},\ \bibinfo {pages} {190401} (\bibinfo {year}
  {2011}{\natexlab{a}})}\BibitemShut {NoStop}%
\bibitem [{\citenamefont {Letokhov}\ and\ \citenamefont
  {Mishin}(1979)}]{Letokhov.1979}%
  \BibitemOpen
  \bibfield  {author} {\bibinfo {author} {\bibfnamefont {V.~S.}\ \bibnamefont
  {Letokhov}}\ and\ \bibinfo {author} {\bibfnamefont {V.~I.}\ \bibnamefont
  {Mishin}},\ }\href {\doibase 10.1016/0030-4018(79)90009-9} {\bibfield
  {journal} {\bibinfo  {journal} {Optics Communications}\ }\textbf {\bibinfo
  {volume} {29}},\ \bibinfo {pages} {168} (\bibinfo {year} {1979})}\BibitemShut
  {NoStop}%
\bibitem [{\citenamefont {Fedosseev}\ \emph {et~al.}(2012)\citenamefont
  {Fedosseev}, \citenamefont {Kudryavtsev},\ and\ \citenamefont
  {Mishin}}]{Fedosseev.2012}%
  \BibitemOpen
  \bibfield  {author} {\bibinfo {author} {\bibfnamefont {V.~N.}\ \bibnamefont
  {Fedosseev}}, \bibinfo {author} {\bibfnamefont {Y.}~\bibnamefont
  {Kudryavtsev}}, \ and\ \bibinfo {author} {\bibfnamefont {V.~I.}\ \bibnamefont
  {Mishin}},\ }\href {\doibase \url{10.1088/0031-8949/85/05/058104}} {\bibfield
   {journal} {\bibinfo  {journal} {{Physica Scripta}}\ }\textbf {\bibinfo
  {volume} {85}},\ \bibinfo {pages} {058104} (\bibinfo {year}
  {2012})}\BibitemShut {NoStop}%
\bibitem [{\citenamefont {Lassen}\ \emph {et~al.}(2006)\citenamefont {Lassen},
  \citenamefont {Bricault}, \citenamefont {Dombsky}, \citenamefont {Lavoie},
  \citenamefont {Geppert},\ and\ \citenamefont {Wendt}}]{Lassen.2006}%
  \BibitemOpen
  \bibfield  {author} {\bibinfo {author} {\bibfnamefont {J.}~\bibnamefont
  {Lassen}}, \bibinfo {author} {\bibfnamefont {P.}~\bibnamefont {Bricault}},
  \bibinfo {author} {\bibfnamefont {M.}~\bibnamefont {Dombsky}}, \bibinfo
  {author} {\bibfnamefont {J.~P.}\ \bibnamefont {Lavoie}}, \bibinfo {author}
  {\bibfnamefont {C.}~\bibnamefont {Geppert}}, \ and\ \bibinfo {author}
  {\bibfnamefont {K.}~\bibnamefont {Wendt}},\ }\href {\doibase
  \url{10.1007/s10751-005-9212-2}} {\bibfield  {journal} {\bibinfo  {journal}
  {{Hyperfine Interactions}}\ }\textbf {\bibinfo {volume} {162}},\ \bibinfo
  {pages} {69} (\bibinfo {year} {2006})}\BibitemShut {NoStop}%
\bibitem [{\citenamefont {Cocolios}\ \emph {et~al.}(2013)\citenamefont
  {Cocolios}, \citenamefont {{Al Suradi}}, \citenamefont {Billowes},
  \citenamefont {Budin{\v{c}}evi{\'c}}, \citenamefont {de~Groote},
  \citenamefont {de~Schepper}, \citenamefont {Fedosseev}, \citenamefont
  {Flanagan}, \citenamefont {Franchoo}, \citenamefont {{Garcia Ruiz}},
  \citenamefont {Heylen}, \citenamefont {{Le Blanc}}, \citenamefont {Lynch},
  \citenamefont {Marsh}, \citenamefont {Mason}, \citenamefont {Neyens},
  \citenamefont {Papuga}, \citenamefont {Procter}, \citenamefont {Rajabali},
  \citenamefont {Rossel}, \citenamefont {Rothe}, \citenamefont {Simpson},
  \citenamefont {Smith}, \citenamefont {Strashnov}, \citenamefont {Stroke},
  \citenamefont {Verney}, \citenamefont {Walker}, \citenamefont {Wendt},\ and\
  \citenamefont {Wood}}]{Cocolios.2013}%
  \BibitemOpen
  \bibfield  {author} {\bibinfo {author} {\bibfnamefont {T.~E.}\ \bibnamefont
  {Cocolios}}, \bibinfo {author} {\bibfnamefont {H.~H.}\ \bibnamefont {{Al
  Suradi}}}, \bibinfo {author} {\bibfnamefont {J.}~\bibnamefont {Billowes}},
  \bibinfo {author} {\bibfnamefont {I.}~\bibnamefont {Budin{\v{c}}evi{\'c}}},
  \bibinfo {author} {\bibfnamefont {R.~P.}\ \bibnamefont {de~Groote}}, \bibinfo
  {author} {\bibfnamefont {S.}~\bibnamefont {de~Schepper}}, \bibinfo {author}
  {\bibfnamefont {V.~N.}\ \bibnamefont {Fedosseev}}, \bibinfo {author}
  {\bibfnamefont {K.~T.}\ \bibnamefont {Flanagan}}, \bibinfo {author}
  {\bibfnamefont {S.}~\bibnamefont {Franchoo}}, \bibinfo {author}
  {\bibfnamefont {R.~F.}\ \bibnamefont {{Garcia Ruiz}}}, \bibinfo {author}
  {\bibfnamefont {H.}~\bibnamefont {Heylen}}, \bibinfo {author} {\bibfnamefont
  {F.}~\bibnamefont {{Le Blanc}}}, \bibinfo {author} {\bibfnamefont {K.~M.}\
  \bibnamefont {Lynch}}, \bibinfo {author} {\bibfnamefont {B.~A.}\ \bibnamefont
  {Marsh}}, \bibinfo {author} {\bibfnamefont {P.}~\bibnamefont {Mason}},
  \bibinfo {author} {\bibfnamefont {G.}~\bibnamefont {Neyens}}, \bibinfo
  {author} {\bibfnamefont {J.}~\bibnamefont {Papuga}}, \bibinfo {author}
  {\bibfnamefont {T.~J.}\ \bibnamefont {Procter}}, \bibinfo {author}
  {\bibfnamefont {M.~M.}\ \bibnamefont {Rajabali}}, \bibinfo {author}
  {\bibfnamefont {R.~E.}\ \bibnamefont {Rossel}}, \bibinfo {author}
  {\bibfnamefont {S.}~\bibnamefont {Rothe}}, \bibinfo {author} {\bibfnamefont
  {G.~S.}\ \bibnamefont {Simpson}}, \bibinfo {author} {\bibfnamefont {A.~J.}\
  \bibnamefont {Smith}}, \bibinfo {author} {\bibfnamefont {I.}~\bibnamefont
  {Strashnov}}, \bibinfo {author} {\bibfnamefont {H.~H.}\ \bibnamefont
  {Stroke}}, \bibinfo {author} {\bibfnamefont {D.}~\bibnamefont {Verney}},
  \bibinfo {author} {\bibfnamefont {P.~M.}\ \bibnamefont {Walker}}, \bibinfo
  {author} {\bibfnamefont {K.}~\bibnamefont {Wendt}}, \ and\ \bibinfo {author}
  {\bibfnamefont {R.~T.}\ \bibnamefont {Wood}},\ }\href {\doibase
  \url{10.1016/j.nimb.2013.05.088}} {\bibfield  {journal} {\bibinfo  {journal}
  {{Nuclear Instruments and Methods in Physics Research Section B: Beam
  Interactions with Materials and Atoms}}\ }\textbf {\bibinfo {volume} {317}},\
  \bibinfo {pages} {565} (\bibinfo {year} {2013})}\BibitemShut {NoStop}%
\bibitem [{\citenamefont {Rothe}\ \emph {et~al.}(2013)\citenamefont {Rothe},
  \citenamefont {Andreyev}, \citenamefont {Antalic}, \citenamefont
  {Borschevsky}, \citenamefont {Capponi}, \citenamefont {Cocolios},
  \citenamefont {de~Witte}, \citenamefont {Eliav}, \citenamefont {Fedorov},
  \citenamefont {Fedosseev}, \citenamefont {Fink}, \citenamefont {Fritzsche},
  \citenamefont {Ghys}, \citenamefont {Huyse}, \citenamefont {Imai},
  \citenamefont {Kaldor}, \citenamefont {Kudryavtsev}, \citenamefont
  {K{\"o}ster}, \citenamefont {Lane}, \citenamefont {Lassen}, \citenamefont
  {Liberati}, \citenamefont {Lynch}, \citenamefont {Marsh}, \citenamefont
  {Nishio}, \citenamefont {Pauwels}, \citenamefont {Pershina}, \citenamefont
  {Popescu}, \citenamefont {Procter}, \citenamefont {Radulov}, \citenamefont
  {Raeder}, \citenamefont {Rajabali}, \citenamefont {Rapisarda}, \citenamefont
  {Rossel}, \citenamefont {Sandhu}, \citenamefont {Seliverstov}, \citenamefont
  {Sj{\"o}din}, \citenamefont {{van den Bergh}}, \citenamefont {{van Duppen}},
  \citenamefont {Venhart}, \citenamefont {Wakabayashi},\ and\ \citenamefont
  {Wendt}}]{Rothe.2013}%
  \BibitemOpen
  \bibfield  {author} {\bibinfo {author} {\bibfnamefont {S.}~\bibnamefont
  {Rothe}}, \bibinfo {author} {\bibfnamefont {A.~N.}\ \bibnamefont {Andreyev}},
  \bibinfo {author} {\bibfnamefont {S.}~\bibnamefont {Antalic}}, \bibinfo
  {author} {\bibfnamefont {A.}~\bibnamefont {Borschevsky}}, \bibinfo {author}
  {\bibfnamefont {L.}~\bibnamefont {Capponi}}, \bibinfo {author} {\bibfnamefont
  {T.~E.}\ \bibnamefont {Cocolios}}, \bibinfo {author} {\bibfnamefont
  {H.}~\bibnamefont {de~Witte}}, \bibinfo {author} {\bibfnamefont
  {E.}~\bibnamefont {Eliav}}, \bibinfo {author} {\bibfnamefont {D.~V.}\
  \bibnamefont {Fedorov}}, \bibinfo {author} {\bibfnamefont {V.~N.}\
  \bibnamefont {Fedosseev}}, \bibinfo {author} {\bibfnamefont {D.~A.}\
  \bibnamefont {Fink}}, \bibinfo {author} {\bibfnamefont {S.}~\bibnamefont
  {Fritzsche}}, \bibinfo {author} {\bibfnamefont {L.}~\bibnamefont {Ghys}},
  \bibinfo {author} {\bibfnamefont {M.}~\bibnamefont {Huyse}}, \bibinfo
  {author} {\bibfnamefont {N.}~\bibnamefont {Imai}}, \bibinfo {author}
  {\bibfnamefont {U.}~\bibnamefont {Kaldor}}, \bibinfo {author} {\bibfnamefont
  {Y.}~\bibnamefont {Kudryavtsev}}, \bibinfo {author} {\bibfnamefont
  {U.}~\bibnamefont {K{\"o}ster}}, \bibinfo {author} {\bibfnamefont {J.~F.~W.}\
  \bibnamefont {Lane}}, \bibinfo {author} {\bibfnamefont {J.}~\bibnamefont
  {Lassen}}, \bibinfo {author} {\bibfnamefont {V.}~\bibnamefont {Liberati}},
  \bibinfo {author} {\bibfnamefont {K.~M.}\ \bibnamefont {Lynch}}, \bibinfo
  {author} {\bibfnamefont {B.~A.}\ \bibnamefont {Marsh}}, \bibinfo {author}
  {\bibfnamefont {K.}~\bibnamefont {Nishio}}, \bibinfo {author} {\bibfnamefont
  {D.}~\bibnamefont {Pauwels}}, \bibinfo {author} {\bibfnamefont
  {V.}~\bibnamefont {Pershina}}, \bibinfo {author} {\bibfnamefont
  {L.}~\bibnamefont {Popescu}}, \bibinfo {author} {\bibfnamefont {T.~J.}\
  \bibnamefont {Procter}}, \bibinfo {author} {\bibfnamefont {D.}~\bibnamefont
  {Radulov}}, \bibinfo {author} {\bibfnamefont {S.}~\bibnamefont {Raeder}},
  \bibinfo {author} {\bibfnamefont {M.~M.}\ \bibnamefont {Rajabali}}, \bibinfo
  {author} {\bibfnamefont {E.}~\bibnamefont {Rapisarda}}, \bibinfo {author}
  {\bibfnamefont {R.~E.}\ \bibnamefont {Rossel}}, \bibinfo {author}
  {\bibfnamefont {K.}~\bibnamefont {Sandhu}}, \bibinfo {author} {\bibfnamefont
  {M.~D.}\ \bibnamefont {Seliverstov}}, \bibinfo {author} {\bibfnamefont
  {A.~M.}\ \bibnamefont {Sj{\"o}din}}, \bibinfo {author} {\bibfnamefont
  {P.}~\bibnamefont {{van den Bergh}}}, \bibinfo {author} {\bibfnamefont
  {P.}~\bibnamefont {{van Duppen}}}, \bibinfo {author} {\bibfnamefont
  {M.}~\bibnamefont {Venhart}}, \bibinfo {author} {\bibfnamefont
  {Y.}~\bibnamefont {Wakabayashi}}, \ and\ \bibinfo {author} {\bibfnamefont
  {K.~D.~A.}\ \bibnamefont {Wendt}},\ }\href {\doibase
  \url{10.1038/ncomms2819}} {\bibfield  {journal} {\bibinfo  {journal} {{Nature
  communications}}\ }\textbf {\bibinfo {volume} {4}},\ \bibinfo {pages} {1835}
  (\bibinfo {year} {2013})}\BibitemShut {NoStop}%
\bibitem [{\citenamefont {de~Groote}\ \emph {et~al.}(2017)\citenamefont
  {de~Groote}, \citenamefont {Verlinde}, \citenamefont {Sonnenschein},
  \citenamefont {Flanagan}, \citenamefont {Moore},\ and\ \citenamefont
  {Neyens}}]{Groote.2017}%
  \BibitemOpen
  \bibfield  {author} {\bibinfo {author} {\bibfnamefont {R.~P.}\ \bibnamefont
  {de~Groote}}, \bibinfo {author} {\bibfnamefont {M.}~\bibnamefont {Verlinde}},
  \bibinfo {author} {\bibfnamefont {V.}~\bibnamefont {Sonnenschein}}, \bibinfo
  {author} {\bibfnamefont {K.~T.}\ \bibnamefont {Flanagan}}, \bibinfo {author}
  {\bibfnamefont {I.}~\bibnamefont {Moore}}, \ and\ \bibinfo {author}
  {\bibfnamefont {G.}~\bibnamefont {Neyens}},\ }\href {\doibase
  \url{10.1103/PhysRevA.95.032502}} {\bibfield  {journal} {\bibinfo  {journal}
  {{Physical Review A}}\ }\textbf {\bibinfo {volume} {95}},\ \bibinfo {pages}
  {1693} (\bibinfo {year} {2017})}\BibitemShut {NoStop}%
\bibitem [{\citenamefont {Sonnenschein}\ \emph {et~al.}(2012)\citenamefont
  {Sonnenschein}, \citenamefont {Raeder}, \citenamefont {Hakimi}, \citenamefont
  {Moore},\ and\ \citenamefont {Wendt}}]{Sonnenschein.2012}%
  \BibitemOpen
  \bibfield  {author} {\bibinfo {author} {\bibfnamefont {V.}~\bibnamefont
  {Sonnenschein}}, \bibinfo {author} {\bibfnamefont {S.}~\bibnamefont
  {Raeder}}, \bibinfo {author} {\bibfnamefont {A.}~\bibnamefont {Hakimi}},
  \bibinfo {author} {\bibfnamefont {I.~D.}\ \bibnamefont {Moore}}, \ and\
  \bibinfo {author} {\bibfnamefont {K.}~\bibnamefont {Wendt}},\ }\href
  {\doibase \url{10.1088/0953-4075/45/16/165005}} {\bibfield  {journal}
  {\bibinfo  {journal} {{Journal of Physics B: Atomic, Molecular and Optical
  Physics}}\ }\textbf {\bibinfo {volume} {45}},\ \bibinfo {pages} {165005}
  (\bibinfo {year} {2012})}\BibitemShut {NoStop}%
\bibitem [{\citenamefont {Rothe}\ \emph {et~al.}(2011)\citenamefont {Rothe},
  \citenamefont {Marsh}, \citenamefont {Mattolat}, \citenamefont {Fedosseev},\
  and\ \citenamefont {Wendt}}]{Rothe.2011}%
  \BibitemOpen
  \bibfield  {author} {\bibinfo {author} {\bibfnamefont {S.}~\bibnamefont
  {Rothe}}, \bibinfo {author} {\bibfnamefont {B.~A.}\ \bibnamefont {Marsh}},
  \bibinfo {author} {\bibfnamefont {C.}~\bibnamefont {Mattolat}}, \bibinfo
  {author} {\bibfnamefont {V.~N.}\ \bibnamefont {Fedosseev}}, \ and\ \bibinfo
  {author} {\bibfnamefont {K.}~\bibnamefont {Wendt}},\ }\href {\doibase
  \url{10.1088/1742-6596/312/5/052020}} {\bibfield  {journal} {\bibinfo
  {journal} {{Journal of Physics: Conference Series}}\ }\textbf {\bibinfo
  {volume} {312}},\ \bibinfo {pages} {052020} (\bibinfo {year}
  {2011})}\BibitemShut {NoStop}%
\bibitem [{\citenamefont {Sonnenschein}\ \emph {et~al.}(2014)\citenamefont
  {Sonnenschein}, \citenamefont {Moore}, \citenamefont {Khan}, \citenamefont
  {Pohjalainen},\ and\ \citenamefont {Reponen}}]{Sonnenschein.2014}%
  \BibitemOpen
  \bibfield  {author} {\bibinfo {author} {\bibfnamefont {V.}~\bibnamefont
  {Sonnenschein}}, \bibinfo {author} {\bibfnamefont {I.~D.}\ \bibnamefont
  {Moore}}, \bibinfo {author} {\bibfnamefont {H.}~\bibnamefont {Khan}},
  \bibinfo {author} {\bibfnamefont {I.}~\bibnamefont {Pohjalainen}}, \ and\
  \bibinfo {author} {\bibfnamefont {M.}~\bibnamefont {Reponen}},\ }\href
  {\doibase \url{10.1007/s10751-013-1000-9}} {\bibfield  {journal} {\bibinfo
  {journal} {{Hyperfine Interactions}}\ }\textbf {\bibinfo {volume} {227}},\
  \bibinfo {pages} {113} (\bibinfo {year} {2014})}\BibitemShut {NoStop}%
\bibitem [{\citenamefont {Wolf}\ \emph {et~al.}(2018)\citenamefont {Wolf},
  \citenamefont {Studer}, \citenamefont {Wendt},\ and\ \citenamefont
  {Schmidt-Kaler}}]{Wolf.2018}%
  \BibitemOpen
  \bibfield  {author} {\bibinfo {author} {\bibfnamefont {S.}~\bibnamefont
  {Wolf}}, \bibinfo {author} {\bibfnamefont {D.}~\bibnamefont {Studer}},
  \bibinfo {author} {\bibfnamefont {K.}~\bibnamefont {Wendt}}, \ and\ \bibinfo
  {author} {\bibfnamefont {F.}~\bibnamefont {Schmidt-Kaler}},\ }\href {\doibase
  \url{10.1007/s00340-018-6903-3}} {\bibfield  {journal} {\bibinfo  {journal}
  {{Applied Physics B}}\ }\textbf {\bibinfo {volume} {124}},\ \bibinfo {pages}
  {412} (\bibinfo {year} {2018})}\BibitemShut {NoStop}%
\bibitem [{\citenamefont {Teigelh{\"o}fer}\ \emph {et~al.}(2010)\citenamefont
  {Teigelh{\"o}fer}, \citenamefont {Bricault}, \citenamefont {Chachkova},
  \citenamefont {Gillner}, \citenamefont {Lassen}, \citenamefont {Lavoie},
  \citenamefont {Li}, \citenamefont {Mei{\ss}ner}, \citenamefont {Neu},\ and\
  \citenamefont {Wendt}}]{Teigelhofer.2010}%
  \BibitemOpen
  \bibfield  {author} {\bibinfo {author} {\bibfnamefont {A.}~\bibnamefont
  {Teigelh{\"o}fer}}, \bibinfo {author} {\bibfnamefont {P.}~\bibnamefont
  {Bricault}}, \bibinfo {author} {\bibfnamefont {O.}~\bibnamefont {Chachkova}},
  \bibinfo {author} {\bibfnamefont {M.}~\bibnamefont {Gillner}}, \bibinfo
  {author} {\bibfnamefont {J.}~\bibnamefont {Lassen}}, \bibinfo {author}
  {\bibfnamefont {J.~P.}\ \bibnamefont {Lavoie}}, \bibinfo {author}
  {\bibfnamefont {R.}~\bibnamefont {Li}}, \bibinfo {author} {\bibfnamefont
  {J.}~\bibnamefont {Mei{\ss}ner}}, \bibinfo {author} {\bibfnamefont
  {W.}~\bibnamefont {Neu}}, \ and\ \bibinfo {author} {\bibfnamefont {K.~D.~A.}\
  \bibnamefont {Wendt}},\ }\href {\doibase \url{10.1007/s10751-010-0171-x}}
  {\bibfield  {journal} {\bibinfo  {journal} {{Hyperfine Interactions}}\
  }\textbf {\bibinfo {volume} {196}},\ \bibinfo {pages} {161} (\bibinfo {year}
  {2010})}\BibitemShut {NoStop}%
\bibitem [{\citenamefont {Lu}\ \emph {et~al.}(2011{\natexlab{b}})\citenamefont
  {Lu}, \citenamefont {Youn},\ and\ \citenamefont {Lev}}]{Lu.2011}%
  \BibitemOpen
  \bibfield  {author} {\bibinfo {author} {\bibfnamefont {M.}~\bibnamefont
  {Lu}}, \bibinfo {author} {\bibfnamefont {S.~H.}\ \bibnamefont {Youn}}, \ and\
  \bibinfo {author} {\bibfnamefont {B.~L.}\ \bibnamefont {Lev}},\ }\href
  {\doibase \url{10.1103/PhysRevA.83.012510}} {\bibfield  {journal} {\bibinfo
  {journal} {{Physical Review A}}\ }\textbf {\bibinfo {volume} {83}},\ \bibinfo
  {pages} {012510} (\bibinfo {year} {2011}{\natexlab{b}})}\BibitemShut
  {NoStop}%
\bibitem [{\citenamefont {Studer}\ \emph {et~al.}(2018)\citenamefont {Studer},
  \citenamefont {Maske}, \citenamefont {Windpassinger},\ and\ \citenamefont
  {Wendt}}]{Supplement}%
  \BibitemOpen
  \bibfield  {author} {\bibinfo {author} {\bibfnamefont {D.}~\bibnamefont
  {Studer}}, \bibinfo {author} {\bibfnamefont {L.}~\bibnamefont {Maske}},
  \bibinfo {author} {\bibfnamefont {P.}~\bibnamefont {Windpassinger}}, \ and\
  \bibinfo {author} {\bibfnamefont {K.}~\bibnamefont {Wendt}},\ }\href@noop {}
  {\  (\bibinfo {year} {2018})}\BibitemShut {NoStop}%
\bibitem [{\citenamefont {Studer}\ \emph {et~al.}(2017)\citenamefont {Studer},
  \citenamefont {Dyrauf}, \citenamefont {Naubereit}, \citenamefont {Heinke},\
  and\ \citenamefont {Wendt}}]{Studer.2017}%
  \BibitemOpen
  \bibfield  {author} {\bibinfo {author} {\bibfnamefont {D.}~\bibnamefont
  {Studer}}, \bibinfo {author} {\bibfnamefont {P.}~\bibnamefont {Dyrauf}},
  \bibinfo {author} {\bibfnamefont {P.}~\bibnamefont {Naubereit}}, \bibinfo
  {author} {\bibfnamefont {R.}~\bibnamefont {Heinke}}, \ and\ \bibinfo {author}
  {\bibfnamefont {K.}~\bibnamefont {Wendt}},\ }\href {\doibase
  \url{10.1007/s10751-016-1384-4}} {\bibfield  {journal} {\bibinfo  {journal}
  {{Hyperfine Interactions}}\ }\textbf {\bibinfo {volume} {238}},\ \bibinfo
  {pages} {02A916} (\bibinfo {year} {2017})}\BibitemShut {NoStop}%
\end{thebibliography}%
	
\end{document}